\documentclass[a4paper, amsfonts, amssymb, amsmath, reprint, showkeys, nofootinbib, twoside, superscriptaddress]{revtex4-1}
\usepackage[english]{babel}
\usepackage[utf8]{inputenc}
\usepackage[colorinlistoftodos, color=green!40, prependcaption]{todonotes}
\usepackage[pdftex, pdftitle={Article}, pdfauthor={Author}]{hyperref} 
\usepackage{appendix}
\usepackage{physics}
\usepackage{amsmath}
\usepackage{breqn}
\usepackage{graphicx}
\usepackage{verbatim}
\usepackage[utf8]{inputenc}
\usepackage{color}
\usepackage[normalem]{ulem}
\begin{document}

\title{Error rate reduction of single-qubit gates via noise-aware decomposition into native gates}
\author{Thomas J. Maldonado}
\email{maldonado@princeton.edu}
\affiliation{Department of Electrical and Computer Engineering\(,\) Princeton University\(,\) Princeton\(,\) NJ 08544\(,\) USA}
\affiliation{Department of Physics\(,\) Harvard University\(,\) Cambridge\(,\) MA 02138\(,\) USA}
\author{Johannes Flick}
\affiliation{Center for Computational Quantum Physics\(,\) Flatiron Institute\(,\) 162 Fifth Avenue\(,\) New York\(,\) NY 10010\(,\) USA}
\author{Stefan Krastanov}
\affiliation{Department of Electrical Engineering and Computer Science\(,\) Massachusetts Institute of Technology\(,\) Cambridge\(,\) MA 02139\(,\) USA}
\affiliation{John A. Paulson School of Engineering and Applied Sciences\(,\) Harvard University\(,\) Cambridge\(,\) MA 02138\(,\) USA}
\author{Alexey Galda}
\affiliation{Menten AI\(,\) Inc.\(,\) San Francisco\(,\) CA 94111\(,\) USA}
\affiliation{James Franck Institute\(,\) University of Chicago\(,\) Chicago\(,\) IL 60637\(,\) USA}
\affiliation{Computational Science Division\(,\) Argonne National Laboratory\(,\) Lemont\(,\) IL 60439\(,\) USA}

\begin{abstract}
        In the current era of Noisy Intermediate-Scale Quantum (NISQ) technology, the practical use of quantum computers remains inhibited by our inability to aptly decouple qubits from their environment to mitigate computational errors. In this paper, we introduce an approach by which knowledge of a qubit's initial quantum state and the standard parameters describing its decoherence can be leveraged to mitigate the noise present during the execution of a single-qubit gate. We benchmark our protocol using cloud-based access to IBM quantum processors. On \texttt{ibmq\_rome}, we demonstrate a reduction of the single-qubit error rate by $38$\%, from $1.6 \times 10 ^{-3}$ to $1.0 \times 10 ^{-3}$, provided the initial state of the input qubit is known. On \texttt{ibmq\_bogota}, we prove that our protocol will never decrease gate fidelity, provided the system's \(T_1\) and \(T_2\) times have not drifted above 100 times their assumed values. The protocol can be used to reduce quantum state preparation errors, as well as to improve the fidelity of quantum circuits for which some knowledge of the qubits’ intermediate states can be inferred. This paper presents a pathway to using information about noise levels and quantum state distributions to significantly reduce error rates associated with quantum gates via optimized decomposition into native hardware gates.
\end{abstract}

\flushbottom
\maketitle
\thispagestyle{empty}

\section*{Introduction}
    Four decades after the conception of a quantum computer (QC) \cite{feynman}, its far-reaching computational potential remains abundantly clear \cite{Shor_1997}. Among the various physical systems whose quantum properties can be harnessed for computation  \cite{PhysRevA.57.120, PhysRevLett.74.4091, cooper-pair, silicon}, superconducting transmon qubits have demonstrated promise in their ability to realize scalable QCs \cite{Devoret1169, Wendin_2017, doi:10.1063/1.5089550} and have accordingly been made available to the public via cloud-based services offered by private companies such as IBM, Rigetti Computing, and Amazon. Despite the recent increase in availability, in the current era of Noisy Intermediate-Scale Quantum (NISQ) technology \cite{Preskill2018quantumcomputingin}, our ability to utilize these machines to their full potential remains significantly inhibited by the computational errors that arise from interactions between the physical qubits and their environment.
    
    While these detrimental interactions can be suppressed through the development of noise-resilient quantum hardware, the effective noise present during circuit execution can also be mitigated by optimizations in the compilation process \cite{jurcevic2021demonstration}. Examples of software-based optimization protocols have been demonstrated across the full quantum computing stack, from high-level circuit depth compression via quantum-assisted quantum compiling \cite{Khatri_2019} down to the optimization of individual native gates—the default quantum operations calibrated by the hardware provider—through the use of pulse-level control \cite{gokhale2020optimized, carvalho2020errorrobust, werninghaus2021leakage}. Some notable examples include noise tailoring via randomized compiling \cite{Wallman_2016}, dynamical decoupling of idle qubits \cite{Pokharel_2018}, optimized state preparation via active reset \cite{PhysRevLett.121.060502, PhysRevApplied.10.044030}, and measurement via excited state promoted readout \cite{esp1, PhysRevX.10.011001}.
    
    In this paper, we pioneer a software-based optimization protocol for fidelity improvements of general single-qubit gates by leveraging knowledge of the qubit decoherence parameters to generate an optimized noise-aware decomposition into native hardware gates. The optimization of native gates themselves is a complementary task and a powerful noise-mitigation tool in its own right, but it requires pulse-level control, a level of hardware access both unfamiliar and inaccessible to many users of NISQ devices. The goal of this paper is to demonstrate a reduction of single-qubit error rates without the need for this lower level of access. By optimizing the decomposition of single-qubit gates without improving the native gates themselves, we demonstrate the efficacy of a protocol that is straightforward to implement at the gate level and requires minimal knowledge of the underlying Hamiltonian governing the qubit dynamics during gate execution. Accordingly, it can be easily adapted for use in QCs based on arbitrary physical systems. Our results demonstrate that it is possible to significantly improve the fidelity of single-qubits gates by leveraging knowledge of the qubit's initial state, along with its characteristic coherence times $T_1$ and $T_2$.
    We perform two randomized benchmarking (RB) \cite{PhysRevA.77.012307} experiments: i) on the \texttt{ibmq\_rome} \cite{rome} quantum processor, we empirically determine the reduction of the single-qubit error rate offered by our optimization technique when the initial state of the input qubit is known, and ii) on the \texttt{ibmq\_bogota} \cite{bogota} quantum processor, we analyze the sensitivity of our approach to the accuracy and drifts of the device's calibrated $T_1$ and $T_2$ coherence times. Our results demonstrate that it is possible to reduce the single-qubit error rate by up to $38$\% and that the approach is extremely robust against drifts and miscalibrations of $T_{1,2}$ coherence times, providing measurable fidelity improvements even when the $T_{1,2}$ values are up to 2 orders of magnitude different from the true values, i.e., off by a factor of $0.1$ to $100$.
    While we demonstrate our approach on two five-qubit IBM transmon devices, our optimization protocol is hardware-agnostic and assumes the two most prominent channels of Markovian noise in NISQ devices: relaxation and dephasing.
\section*{Method}
\subsection*{Native Gates} \label{native_gates}
    \begin{figure*}[t]
        \centering
        \includegraphics[width=2.0\columnwidth]{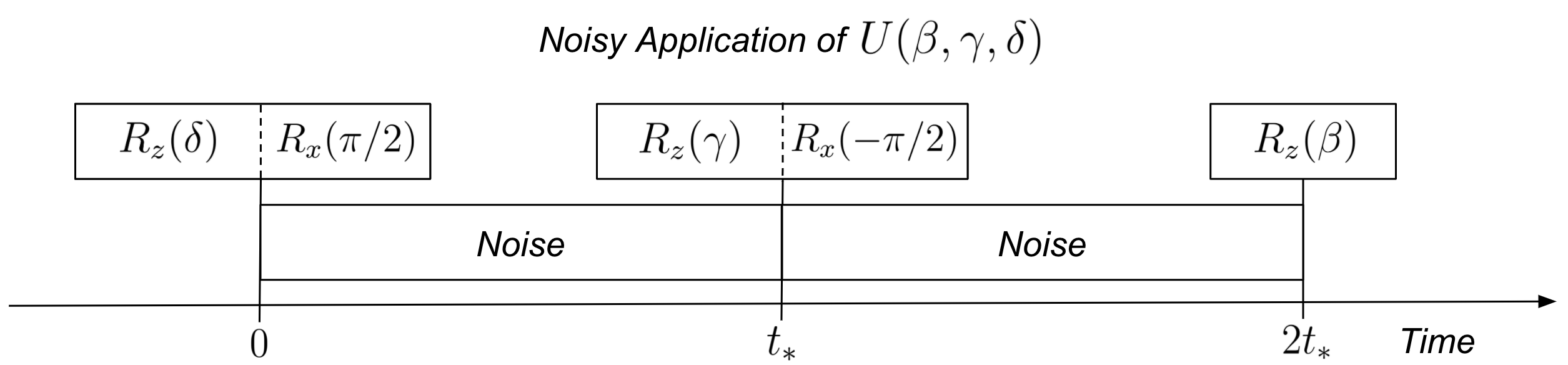}
        \caption{Noisy gate application: a single-qubit gate \(U(\beta, \gamma, \delta)\) is decomposed into native gates according to Eq.~\eqref{eqn:ibm_decomp}. Each \(R_z\) gate is applied via a noiseless frame change. Each \(R_x(\pm\pi/2)\) gate is applied via a microwave pulse and modeled as an instantaneous rotation, followed by decoherence over time \(t_*\) equal to the duration of the microwave pulse.}
        \label{fig:noisy_application}
    \end{figure*}
    We start by introducing IBM's native gate set and describing how the compiler of quantum circuits decomposes single-qubit gates into native gates. Throughout this paper, we use the terms single-qubit gate, rotation, and operation interchangeably. An arbitrary rotation $U$ on the Bloch sphere can be parameterized by its Euler angles \cite{mikeandike}. Concretely, \(\forall U \in SU(2)\), \(\exists \alpha\), \(\beta\), \(\delta \in [0,2\pi)\), \(\gamma \in [0,\pi]\) such that
    \begin{equation}
        \label{eqn:su2}
        U(\beta, \gamma, \delta) = e^{i\alpha} R_z(\beta) R_y(\gamma) R_z(\delta)\,,
    \end{equation}
    where \(R_z\) and \(R_y\) are rotations about the z- and y-axes, respectively. Since a global phase enacted on a quantum state has no physical effect, the value of \(\alpha\) is irrelevant. Effectively, Eq.~\eqref{eqn:su2} defines the decomposition of an arbitrary single-qubit gate into three rotations, two about the z-axis and one about the y-axis. In this paper, we work exclusively with the native gates used by IBM in their standard decomposition framework \cite{cross2017open}: rotations about the x-axis by integer multiples of \(\pi/2\) and rotations about the z-axis by an arbitrary angle. We note that, in principle, IBM's native gate set can be expanded using pulse-level control \cite{gokhale2020optimized, carvalho2020errorrobust}, but in hopes of making our protocol easier to implement for those without this lower level of access, we maintain the standard native gate set described above. The decomposition outlined in Eq.~\eqref{eqn:su2} can be rewritten in terms of these native gates \(R_z\) and \(R_x(\pm\pi/2)\):
    \begin{equation}
        \label{eqn:ibm_decomp}
        U(\beta, \gamma, \delta) = e^{i\alpha} R_z(\beta) R_x(-\pi/2) R_z(\gamma) R_x(\pi/2)R_z(\delta)\,.
    \end{equation}
    Thus, any single-qubit gate can be applied via the sequential application of IBM's native gates defined by Eq.~\eqref{eqn:ibm_decomp}. 
    
    Before proceeding to the noise model, we make a brief note on the physical implementation of the native gate set on IBM quantum processors~\cite{cross2017open}. IBM quantum systems are built using fixed-frequency superconducting transmon qubits, wherein the qubits are manipulated using microwave pulses. The \(R_x(\pm\pi/2)\) gates in the decomposition defined by Eq. ~\eqref{eqn:ibm_decomp} are implemented using calibrated microwave pulses, while the \(R_z\) rotations are realized as zero-duration ``virtual'' gates by adding a phase offset in software~\cite{mckay2017efficient}. For the purposes of the noise model outline below, it is important that microwave pulses of fixed shape and duration are applied to qubits only to implement the \(R_x(\pm\pi/2)\) native gates.
    
\subsection*{Noise Model}\label{noise_model}
    While there are a number of noteworthy sources of single-qubit decoherence, including leakage \cite{werninghaus2021leakage} and non-Markovian noise \cite{RevModPhys.88.021002}, we limit our attention to amplitude damping (relaxation) and dephasing, respectively characterized by IBM's publicly reported \(T_1\) and \(T_2\) coherence times. All single-qubit noise specifications for \texttt{ibmq\_rome} \cite{rome} and \texttt{ibmq\_bogota} \cite{bogota} were provided by IBM through Qiskit \cite{Qiskit} and are tabulated in Appendix \ref{noise_specs}. 
    Based on the physical implementation of IBM's native gates discussed above, we model the noisy application of an \(R_x(\pm\pi/2)\) gate as an instantaneous rotation, followed by decay and dephasing over time \(t_*\) equal to the gate duration. We emphasize that this model is an approximation, and though it is not necessarily exact, it captures enough of the noise dynamics for the purpose of this study. This is mathematically realized via the initial application of an \(R_x(\pm\pi/2)\) unitary, followed by the appropriate Kraus operators. Accordingly, we model the noisy application of a single-qubit gate parameterized by Euler angles \((\beta, \gamma, \delta)\) by applying these Kraus operators after each instance of \(R_x(\pm\pi/2)\) in Eq.~\eqref{eqn:ibm_decomp}; a pictorial representation can be found in Fig. \ref{fig:noisy_application}. In this model, the noisy application of a single-qubit gate with Euler angles \((\beta, \gamma, \delta)\) will transform an initially pure state with Bloch sphere coordinates \((\theta, \phi)\) into a mixed state with the following density matrix: 
     \begin{equation}
        \label{eqn:noisy_density_matrix}
        \rho_{(\beta, \gamma, \delta, \theta, \phi)} = 
        \begin{bmatrix}
            a & b \\
            b^* & 1 - a
        \end{bmatrix}\,,
    \end{equation}
    \begin{align}
        \label{eqn:noisy_density_matrix_a}
        \begin{split}
            a = \frac{1}{2}\bigg{[}&\Big{(}-\sin{\gamma}\cos{(\phi + \delta)}\sin{\theta} + \cos{\gamma}\cos{\theta}\Big{)} \\
            &\times(1 - \lambda\textsubscript{A})^{3/2}\sqrt{1 - \lambda\textsubscript{P}} + 1 + \lambda\textsubscript{A}\bigg{]}
        \end{split}
    \end{align}

    \begin{align}
        \label{eqn:noisy_density_matrix_b}
        \begin{split}
            b = \frac{e^{-i\beta}}{2}\bigg{[}&\Big{(}\cos{(\phi + \delta)}\cos{\gamma}\sin{\theta} 
                 + \sin{\gamma}\cos{\theta}\Big{)} \\
                 &\times(1 - \lambda\textsubscript{A})(1 - \lambda\textsubscript{P}) \\
                 &-i\Big{(}\sin{(\phi + \delta)}\sin{\theta}(1 - \lambda\textsubscript{A}) + \lambda\textsubscript{A}\Big{)} \\
                 &\hspace{0.4cm}\times\sqrt{1 - \lambda\textsubscript{A}}\sqrt{1 - \lambda\textsubscript{P}}\bigg{]}
        \end{split}
    \end{align}
    The derivation of this expression is provided in Appendix \ref{noise_model_derivation}.
    The variable \(\lambda\textsubscript{A}\) is equal to the probability of a spontaneous emission during the application of an \(R_x(\pm\pi/2)\) gate, and \(\lambda\textsubscript{P}\) is equal to the probability of a spontaneous phase flip during the application of an \(R_x(\pm\pi/2)\) gate. Both parameters are defined as functions of the system's \(T_1\) and \(T_2\) times, respectively, along with the \(R_x(\pm\pi/2)\) gate duration \(t_*\):
    \begin{align}
        \label{eqn:lambda_A}
        &\lambda\textsubscript{A}=1-e^{-t_*/{T_1}}\,,\\
        \label{eqn:lambda_P}
        &\lambda\textsubscript{P}=1-e^{-t_*/{T_2}}\,.
    \end{align}
    We note that all time-dependent terms in Eqs.~(\ref{eqn:noisy_density_matrix}--\ref{eqn:noisy_density_matrix_b})~are functions of \(t_*/{T_{1,2}}\).
\subsection*{Optimization} \label{opt_section}
    Using the noise model given by Eqs.~(\ref{eqn:noisy_density_matrix}--\ref{eqn:noisy_density_matrix_b}), we now outline the protocol by which the fidelity of an arbitrary single-qubit gate can be improved. Suppose we wish to implement the target operation \(U(\beta\textsubscript{t}, \gamma\textsubscript{t}, \delta\textsubscript{t})\) acting on the initially pure state with Bloch sphere coordinates \((\theta, \phi)\), represented below by \(\ket{\psi(\theta, \phi)}\). Our protocol amounts to maximizing the following fidelity over the Euler angles \((\beta, \gamma, \delta)\):
    \begin{align}
        \label{eqn:general_fidelity}
        \begin{split}
            &F(\beta\textsubscript{t}, \gamma\textsubscript{t}, \delta\textsubscript{t}, \beta, \gamma, \delta, \theta, \phi) = \\
            &\hspace{0.5cm}\bra{\psi(\theta, \phi)}U(\beta\textsubscript{t}, \gamma\textsubscript{t}, \delta\textsubscript{t})^{\dagger}\rho_{(\beta, \gamma, \delta, \theta, \phi)}U(\beta\textsubscript{t}, \gamma\textsubscript{t}, \delta\textsubscript{t})\ket{\psi(\theta, \phi)} \\
        \end{split}
    \end{align}
    We find the optimal Euler angles \((\beta', \gamma', \delta')\) via gradient descent over the parameters \((\beta, \gamma, \delta)\). In the presence of noise, the native gate decomposition of \(U(\beta', \gamma', \delta')\) will map the initial state \(\ket{\psi(\theta, \phi)}\) to the target state \(U(\beta\textsubscript{t}, \gamma\textsubscript{t}, \delta\textsubscript{t})\ket{\psi(\theta, \phi)}\) with higher fidelity than the default decomposition of \(U(\beta\textsubscript{t}, \gamma\textsubscript{t}, \delta\textsubscript{t})\). We note that Eq.~\eqref{eqn:general_fidelity} has an explicit closed form and that in general, \(U(\beta', \gamma', \delta') \neq U(\beta\textsubscript{t}, \gamma\textsubscript{t}, \delta\textsubscript{t})\), i.e., the optimized decomposition is not constrained to perform the target operation perfectly in the absence of noise. Throughout this study, the gradient descent was performed in Python using the function \texttt{scipy.optimize.minimize} from the SciPy\cite{2020SciPy-NMeth} library with the method \texttt{L-BFGS-B}. For the parameters used in both experiments performed on IBM's hardware, the gradient descent to optimize a single gate could be performed in approximately 0.2 seconds on a standard computer.
    
\begin{figure}[t]
        \centering
        \includegraphics[width=1.0\columnwidth]{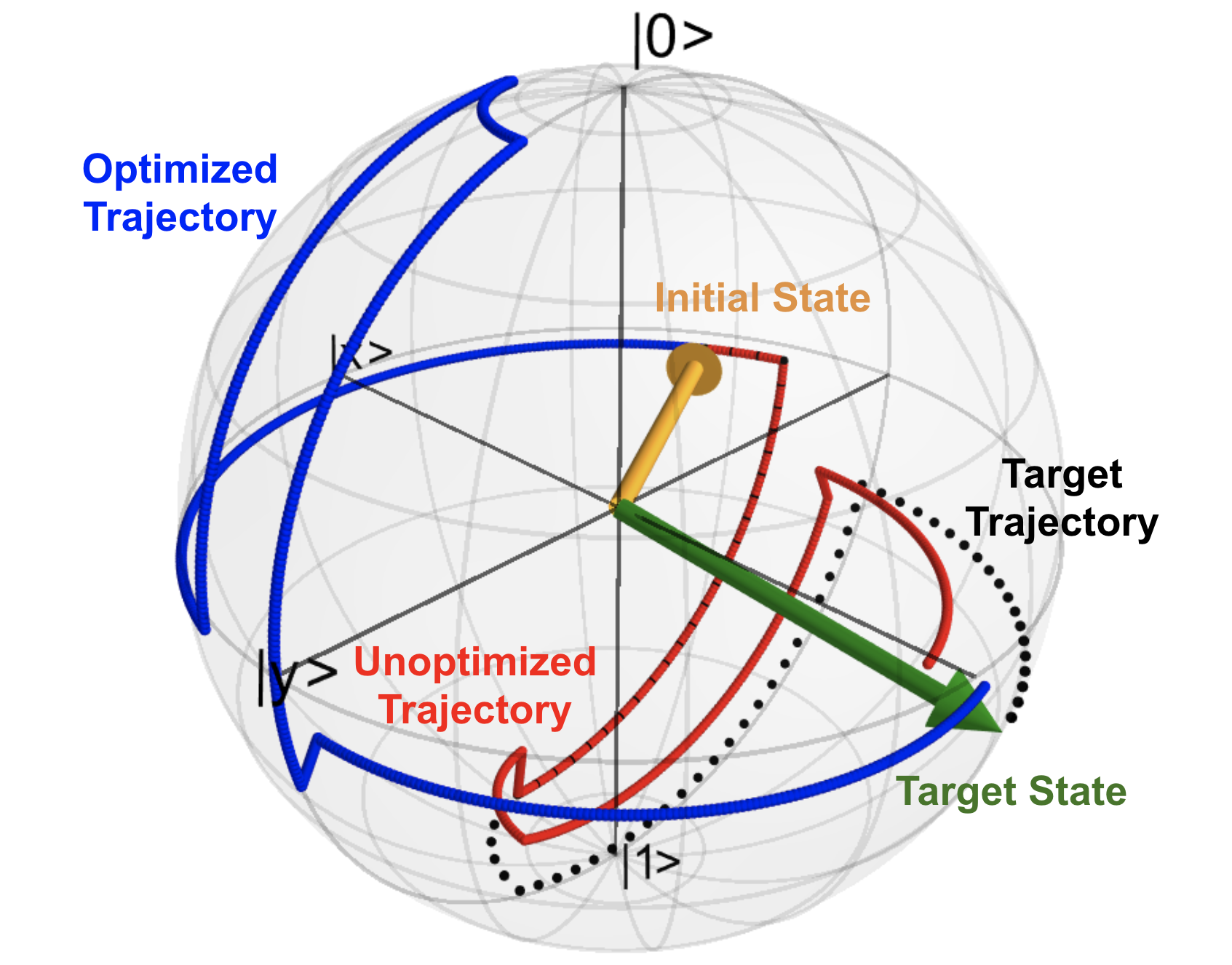}
        \caption{Optimized vs. unoptimized trajectories: an optimized trajectory (blue) and unoptimized trajectory (red) are mapped through the Bloch sphere in the presence of simulated noise corresponding to coherence times 100x shorter than the coherence times of \texttt{ibmq\_rome} qubit 3 on its date of use (07/14/20). Also depicted is the initial state (orange), target trajectory (dotted black), and target state (green). To maximize fidelity, the optimized trajectory evolves the qubit through an intermediate state (after the \(R_x(\pi/2)\) pulse) that avoids regions of the Bloch sphere (south pole and equator) that are more susceptible to noise.}
        \label{fig:bloch_sphere}
    \end{figure}

    To provide some intuition for how the optimized operation improves the fidelity, we begin by noting that the effect of amplitude damping is most pronounced on the south pole of the Bloch sphere (excited state), and the effect of phase damping is most pronounced on the equator of the Bloch sphere (equal superposition states). Thus, to best map the initial state to the target state, the optimizer finds a trajectory through the Bloch sphere that most aptly avoids these noisy regions. For the sake of visualization, we have included an example of an optimized (blue) and an  unoptimized (red) trajectory through the Bloch sphere in Fig. \ref{fig:bloch_sphere}, wherein the optimized trajectory tends towards the coherent north pole more than its unoptimized counterpart. Fig. \ref{fig:bloch_sphere} was generated using QuTiP \cite{qutip1, qutip2}.
    
\section*{Results}\label{results}

\subsection*{Error Rate Reduction on \texttt{ibmq\_rome}}\label{experiment_1}
\begin{figure*}[ht]
    \centering
    \includegraphics[width=2.0\columnwidth]{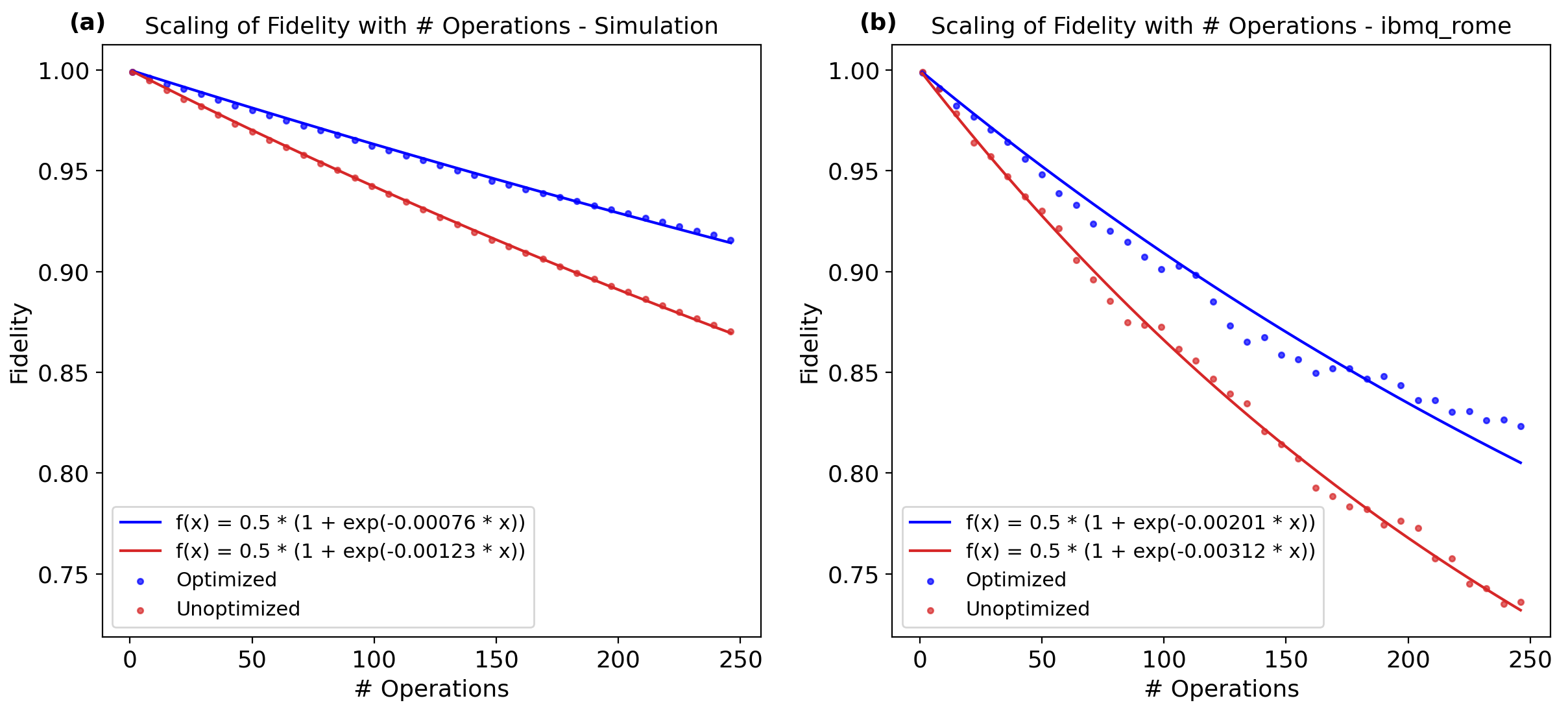}
    \caption{Scaling of fidelity with number of operations: the fidelity (vertical axis) representing the overlap between the state output by the noisy application of each circuit (unoptimized and optimized) and the target state output by the noiseless application of the unoptimized circuit is plotted against the circuit depth (horizontal axis) at which the fidelity was measured. Each data point is the average fidelity of 10 randomized gate sequences with 8,192 shots per measurement. Measurements were taken at circuit depths increasing by 7, \(d \in \{1, 8, 15, ..., 246\}\).}
    \label{fig:experiment_1_plot}
\end{figure*} 
We now experimentally validate the protocol outlined above on qubit 3 of the \texttt{ibmq\_rome} device by empirically determining the reduction in the single-qubit error rate achieved by the optimizer. At the time of the experiment, the qubit had the following characteristics: \(T_1 = 46.4\mu\text{s}\), \(T_2 = 105\mu\text{s}\), pulse duration \(t_* = 35.6\text{ns}\), and therefore damping probabilities \(\lambda\textsubscript{A} = 7.7 \times 10^{-4}\) and \(\lambda\textsubscript{P} = 3.4 \times 10^{-4}\) (see Eqs.~(\ref{eqn:lambda_A}--\ref{eqn:lambda_P})). Because the fidelity improvement offered by the optimizer is relatively small, we use an RB experiment to detect the improvement in fidelity by accumulating it over a long sequence of gates.

We begin by generating a circuit composed of a sequence of randomized single-qubit gates \(C = (G_1,...,G_N)\) acting on the initial state \(\ket{\psi_0} = \ket{0}\). To randomize each gate \(G_i \in C\), we sample its axis of rotation uniformly from the surface of the Bloch sphere and its angle uniformly from the interval \([0,2\pi)\). We note that our method for randomly generating single-qubit gates is one of many and that another popular approach is to sample from the Clifford Group \cite{helsen2019new}. We then optimize each gate \(G_i\) acting on the initial state \(\ket{\psi_{i-1}} = G_{i-1}...G_1\ket{0}\). The result is a new circuit \(C_{\text{opt}}\) composed of the optimized versions of the gates in the circuit \(C\). For each circuit \(C\) and \(C_{\text{opt}}\), we denote the subsequence composed of the first \(d\) gates by \(C^d\) and \(C^d_{\text{opt}}\), respectively. We measure the unoptimized fidelity after \(d\) gates by applying \(C^d\), then applying the 2-pulse native gate decomposition of \((C^d)^{-1}\), and then measuring the probability of collapse to \(\ket{0}\). Similarly, we measure the optimized fidelity by applying \(C^d_{\text{opt}}\), followed by \((C^d)^{-1}\), and then measuring the probability of collapse to \(\ket{0}\). In the presence of noise, the subsequence that maps the \(\ket{0}\) state closer to the target state \(\ket{\psi_d} = C^d\ket{0}\) will return the higher probability of measuring \(\ket{0}\) after \((C^d)^{-1}\) is applied. We generate 10 circuits, each consisting of \(N = 246\) randomized rotations. For each of the 10 circuits, we measure the unoptimized and optimized fidelities at circuit depths increasing by 7, \(d \in \{1,8,15,...,246\}\). The empirically obtained fidelities at each depth \(d\) are then averaged over the 10 circuits. Readout errors were mitigated for all measurements made on \texttt{ibmq\_rome} and \texttt{ibmq\_bogota} by inverting a calibration matrix \cite{readouterror} composed of IBM's publicly reported readout error probabilities.

The results from simulating the execution of the circuits using our noise model and from executing the circuits on \texttt{ibmq\_rome} qubit 3 are displayed in Fig.~\ref{fig:experiment_1_plot}(a) and Fig.~\ref{fig:experiment_1_plot}(b), respectively. In both plots, the blue points represent the average fidelities of the optimized circuits, and the red points represent the average fidelities of the unoptimized circuits. The accumulation of noise in both circuits is reflected in the decrease in fidelity with circuit depth. We fit the data with the ansatz \(f(x) = \frac{1}{2}(1 + \exp{-ax})\) because it satisfies the limiting cases \(f(0) = 1\) and \(\lim_{x\rightarrow\infty}{f(x)}=\frac{1}{2}\). Thus, the error rate of one randomized single-qubit gate is given by the following expression:
\begin{equation}\label{eqn:error_rate}
    \text{error rate} = 1 - \text{fidelity} = 1 - f(1) \approx \frac{a}{2}\,.
\end{equation}
In the simulation and on the hardware, the optimized circuits outperformed the unoptimized circuits, thereby experimentally validating the optimization protocol put forth in this paper. Both the optimized and unoptimized circuits on the hardware have lower fidelities than in the simulation. We attribute this to two factors. First, our noise model only accounts for relaxation and dephasing, when in reality there are other noise channels present, such as sources of non-Markovian noise and leakage. Second, we made the approximation that a noisy single-qubit gate can be modeled as coherent evolution followed by decoherent evolution, when in reality the two occur simultaneously. We believe these two factors are the largest contributors to the lower fidelities observed on the hardware. Finally, from direct calculation of the optimized and unoptimized error rates defined by Eq.~\eqref{eqn:error_rate}, we conclude that on \texttt{ibmq\_rome} qubit 3, our optimization protocol reduces the error rate of a single-qubit gate acting on a known initial state by 38\%, from \(1.6 \times 10^{-3}\) to \(1.0 \times 10^{-3}\). The unoptimized error rate that we report here does not agree with IBM's reported error rate of \(3.4 \times 10^{-4}\). We attribute this discrepancy to differences in methodology when calculating error rates, as well as to other sources of error not included in the model, such as coherent or calibration errors. The discrepancy is also likely attributable to a bias introduced by the 10 randomized gate sequences used in the experiment\cite{2018}. Nonetheless, we maintain our reported error rate reduction as an estimate for the degree to which the noise present during the execution of a single-qubit gate can be mitigated by leveraging knowledge of the initial state of the input qubit and a description of the noise present during gate execution. An experiment of identical structure was carried out on Rigetti's Aspen-8~\cite{rigetti} device and is detailed in Appendix \ref{experiment_3}.

\subsection*{Single-Qubit State Preparation}\label{prep}
    As an illustrative example application of the optimization protocol outlined above, we now analyze its ability to improve the preparation fidelity of a single-qubit state. Suppose that we wish to implement the target operation \(U(\phi\textsubscript{t}, \theta\textsubscript{t}, 0)\) mapping the initial state \(\ket{0}\) to the target state with Bloch sphere coordinates \((\theta\textsubscript{t}, \phi\textsubscript{t})\): 
    \begin{equation}
        \ket{\psi(\theta\textsubscript{t}, \phi\textsubscript{t})} = U(\phi\textsubscript{t}, \theta\textsubscript{t}, 0)\ket{0}\,.
    \end{equation}
    Without loss of generality, we set \(\delta = \theta = \phi = 0\) and reduce Eq.~\eqref{eqn:general_fidelity} to the following:
    \begin{equation}
        \label{eqn:prep_fidelity_1}
        F(\theta\textsubscript{t}, \phi\textsubscript{t}, \beta, \gamma) = \bra{\psi(\theta\textsubscript{t}, \phi\textsubscript{t})}\rho_{(\beta, \gamma, 0,0,0)}\ket{\psi(\theta\textsubscript{t}, \phi\textsubscript{t})}\,.
    \end{equation}
    We find the optimal Euler angles \((\beta', \gamma', 0)\) via gradient descent over the parameters \(\beta\) and \(\gamma\). In the presence of noise, the native gate decomposition of \(U(\beta', \gamma', 0')\) will map the initial state \(\ket{0}\) to the target state \(\ket{\psi(\theta\textsubscript{t}, \phi\textsubscript{t})}\) with higher fidelity than the default decomposition of \(U(\phi\textsubscript{t}, \theta\textsubscript{t}, 0)\). 
    
    We now proceed by analyzing the relationship between the improvement in preparation fidelity offered by the optimizer and the amount of noise in the system. Since \(\lambda\textsubscript{A}\) and \(\lambda\textsubscript{P}\) are typically of comparable magnitude, we consider the improvement offered by optimization in the presence of noise parameterized by \(\lambda = \lambda\textsubscript{A} = \lambda\textsubscript{P}\). For a fixed noise level \(\lambda\), we randomly sample the target state \(\ket{\psi(\theta\textsubscript{t}, \phi\textsubscript{t})}\) uniformly from the surface of the Bloch sphere. We then find the optimal angles \(\beta'\) and \(\gamma'\) and  simulate the application of  \(U(\phi\textsubscript{t}, \theta\textsubscript{t}, 0)\) and \(U(\beta', \gamma', 0)\) on the input \(\ket{0}\). Finally, we calculate the increase in fidelity to the target state \(\ket{\psi(\theta\textsubscript{t}, \phi\textsubscript{t})}\) and repeat this 100 times to find the average increase in preparation fidelity from optimization. The results from simulating the state preparation  over a range of possible noise levels \(\lambda\) are displayed in Fig. \ref{fig:avg_prep_fidelity_plot}.

\subsection*{Knowledge of the Initial State}\label{knowledge}
    The above optimization protocol requires knowledge of the initial state to achieve an improvement in fidelity, as reflected in the fidelity function's explicit dependence on the initial state's Bloch sphere coordinates (see Eq.~\eqref{eqn:general_fidelity}). However, perfect knowledge of the initial state is not necessarily required for optimization. Provided a probability density function \(p(\theta,\phi)\) for the distribution of the initial state over the Bloch sphere, we can optimize the target operation \(U(\beta\textsubscript{t}, \gamma\textsubscript{t}, \delta\textsubscript{t})\) by maximizing the expected fidelity:
    \begin{align}
        \label{eqn:expected_fidelity}
        \begin{split}
            &\langle F(\beta\textsubscript{t}, \gamma\textsubscript{t}, \delta\textsubscript{t}, \beta, \gamma, \delta)\rangle = \\
            &\hspace{0.5cm}\int_0^{2\pi}\int_0^{\pi}F(\beta\textsubscript{t}, \gamma\textsubscript{t}, \delta\textsubscript{t}, \beta, \gamma, \delta, \theta, \phi)p(\theta, \phi)d\theta d\phi
        \end{split}
    \end{align}
    The fidelity function \(F(\beta\textsubscript{t}, \gamma\textsubscript{t}, \delta\textsubscript{t}, \beta, \gamma, \delta, \theta, \phi)\) is defined in Eq.~\eqref{eqn:general_fidelity}. We find the optimal Euler angles \((\beta', \gamma', \delta')\) via gradient descent over the parameters \((\beta, \gamma, \delta)\). In the presence of noise, the native gate decomposition of \(U(\beta', \gamma', \delta')\) will on average map an initial state \(\ket{\psi(\theta, \phi)}\) sampled from the distribution defined by \(p(\theta, \phi)\) to the target state \(U(\beta\textsubscript{t}, \gamma\textsubscript{t}, \delta\textsubscript{t})\ket{\psi(\theta, \phi)}\) with higher fidelity than the default decomposition of \(U(\beta\textsubscript{t}, \gamma\textsubscript{t}, \delta\textsubscript{t})\). \begin{figure}[t]
        \centering
        \includegraphics[width=1.0\columnwidth]{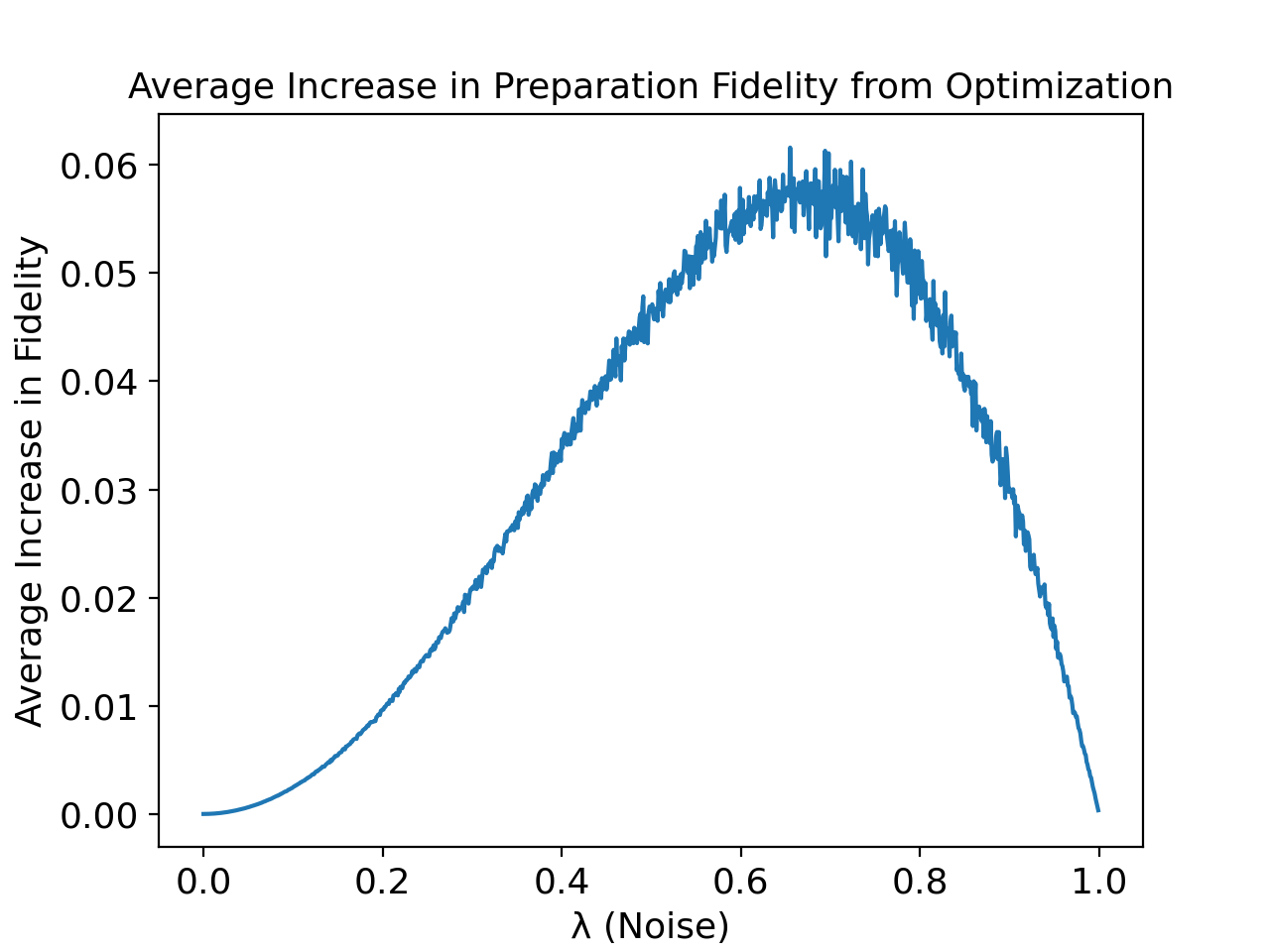}
        \caption{Average increase in preparation fidelity from optimization: the average increase in preparation fidelity (vertical axis) is plotted against the simulated noise level \(\lambda\) (horizontal axis). For each of the 1,000 evenly spaced noise levels, the increase in fidelity is averaged from 100 randomly sampled target states.}
        \label{fig:avg_prep_fidelity_plot}
    \end{figure}
    
    We proceed by analyzing the expected fidelity improvement offered by our protocol as a function of the initial state uncertainty. We have already shown a nontrivial improvement in expected fidelity provided perfect knowledge (i.e., minimal uncertainty) of the initial state \(\ket{0}\) in Fig. \ref{fig:avg_prep_fidelity_plot}. The other extreme corresponds to no knowledge (i.e., maximal uncertainty) of the initial state and is represented by the distribution in which all states are equally likely:
    \begin{equation}
        p(\theta, \phi) = \frac{1}{4\pi}\sin{\theta}\,.
    \end{equation}
    The expected fidelity in this case is given by
    \begin{align}\label{eq:uniform}
        \begin{split}
            &\langle F(\beta\textsubscript{t}, \gamma\textsubscript{t}, \delta\textsubscript{t}, \beta, \gamma, \delta)\rangle = \\ &\hspace{0.5cm}\frac{1}{4\pi}\int_0^{2\pi}\int_0^{\pi}F(\beta\textsubscript{t}, \gamma\textsubscript{t}, \delta\textsubscript{t}, \beta, \gamma, \delta, \theta, \phi)\sin{\theta}d\theta d\phi
        \end{split}
    \end{align}
    Further analysis of Eq. \eqref{eq:uniform} shows
    \begin{equation}
        \nabla_{(\beta, \gamma, \delta)}{\langle F \rangle}\Bigr\rvert_{(\beta, \gamma, \delta) = (\beta\textsubscript{t}, \gamma\textsubscript{t}, \delta\textsubscript{t})} = 0\,.
    \end{equation}

    Regardless of the target operation and the amount of noise, the expected fidelity achieves a local maximum at ${(\beta, \gamma, \delta) = (\beta\textsubscript{t}, \gamma\textsubscript{t}, \delta\textsubscript{t})}$. We deduce that our protocol requires \emph{some} knowledge of the initial state to improve expected fidelity. To further analyze this dependence, we examine the effect of maximizing the expected fidelity for a new probability density function:
    
    \begin{figure}[t]
        \centering
        \includegraphics[width=1.0\columnwidth]{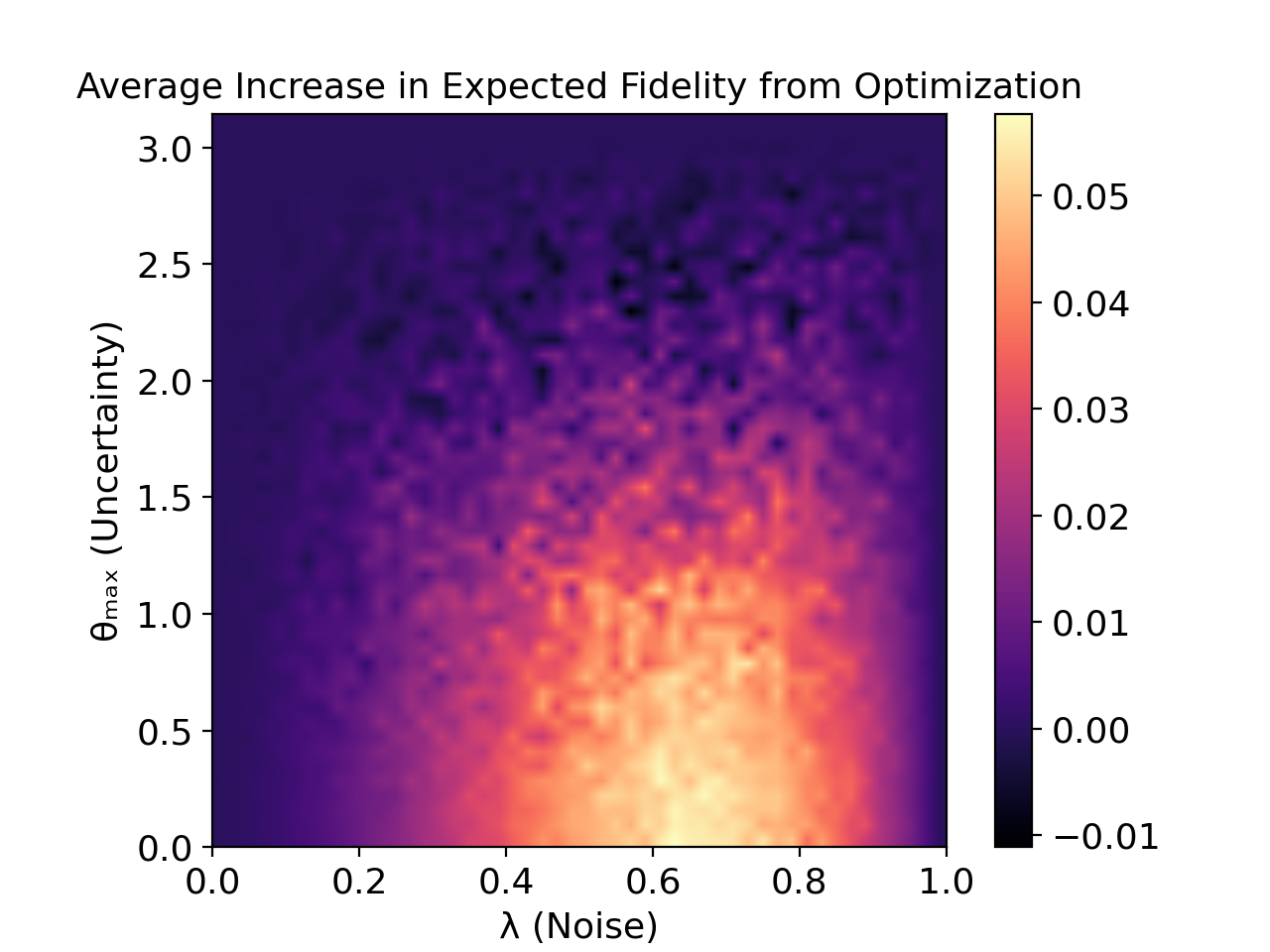}
        \caption{Average increase in expected fidelity from optimization: the average increase in expected fidelity (color bar) is plotted against the simulated noise level \(\lambda\) (horizontal axis) and the uncertainty in the initial state \(\theta\textsubscript{max}\) (vertical axis). $\theta\textsubscript{max} = 0$ corresponds to perfect knowledge of the initial state $\ket{0}$ and therefore recovery of Fig. \ref{fig:avg_prep_fidelity_plot}; \(\theta\textsubscript{max} = \pi\) corresponds to no knowledge of the initial state and therefore no fidelity improvement. Each of the 50x50 data points was generated from 100 randomly sampled target operations and initial states.}
        \label{fig:knowledge_noise_plot}
    \end{figure}
    \begin{figure*}[ht]
        \centering
       \includegraphics[width=2.0\columnwidth]{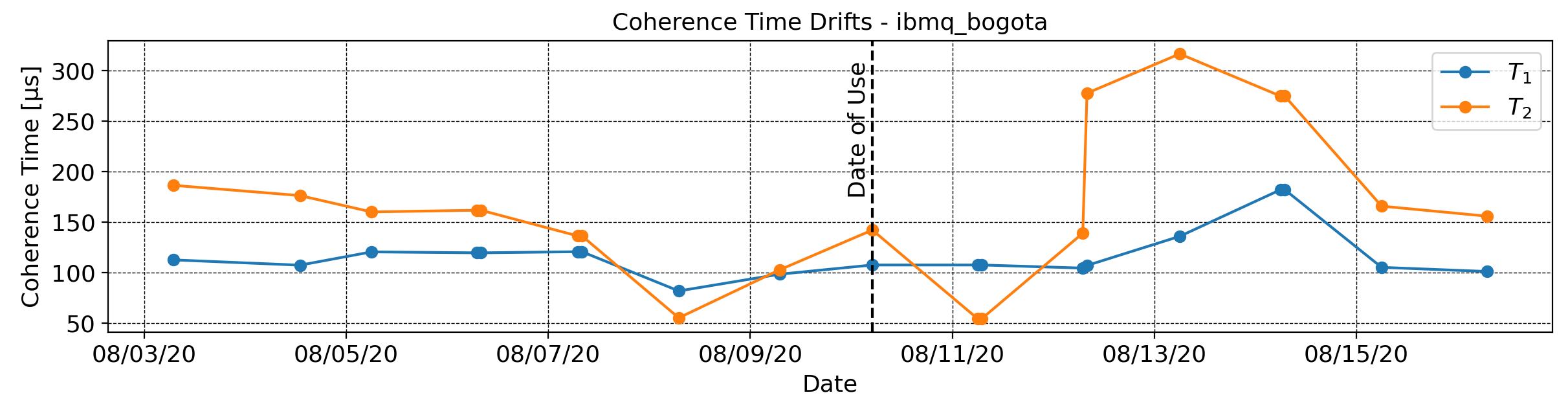}
        \caption{
        Coherence time drifts on \texttt{ibmq\_bogota} qubit 2: IBM's publicly available coherence times on \texttt{ibmq\_bogota} qubit 2 (vertical axis) are plotted against the date of calibration (horiztonal axis) up to one week before and after it was used in this study (08/10/20). Some of the reported coherence times violate \(T_2 \leq 2 \cdot T_1\), thereby suggesting calibration errors.}
        \label{fig:bogota_drifts}
    \end{figure*}  
    \begin{equation}
        p(\theta, \phi) = 
        \begin{cases}
            \frac{\sin{\theta}}{2\pi(1-\cos{\theta\textsubscript{max}})} &\mbox{if } \theta < \theta\textsubscript{max}\,, \\
            \hspace{1cm}0 & \mbox{if } \theta \geq \theta\textsubscript{max}\,.
        \end{cases}
    \end{equation}
    The initial state is now uniformly distributed over the portion of the Bloch sphere with polar angle less than \(\theta\textsubscript{max}\). Accordingly, as \(\theta\textsubscript{max}\) approaches 0, we recover the case of state preparation, and for \(\theta\textsubscript{max} = \pi\), we recover the case of maximal uncertainty. By varying \(\theta\textsubscript{max}\) and examining its effect on the improvement in expected fidelity achieved by maximizing
    \begin{align}
        \label{eqn:EF_theta_max}
        \begin{split}
            &\langle F(\beta\textsubscript{t}, \gamma\textsubscript{t}, \delta\textsubscript{t}, \beta, \gamma, \delta)\rangle = \\ &\hspace{0.3cm}\int_0^{2\pi}\int_0^{\theta\textsubscript{max}}\frac{F(\beta\textsubscript{t}, \gamma\textsubscript{t}, \delta\textsubscript{t}, \beta, \gamma, \delta, \theta, \phi)\sin{\theta}}{2\pi(1-\cos{\theta\textsubscript{max}})}d\theta d\phi
        \end{split}
    \end{align}
    we quantify the degree to which an arbitrary single-qubit gate can be optimized as a function of the initial state uncertainty. To visualize this dependence, we begin by fixing the noise parameter \(\lambda = \lambda\textsubscript{A} = \lambda\textsubscript{P}\) and the maximum polar angle \(\theta\textsubscript{max}\). We proceed to randomly generate a target rotation \(U(\beta\textsubscript{t}, \gamma\textsubscript{t}, \delta\textsubscript{t})\) by sampling the axis of rotation uniformly from the surface of the Bloch sphere and sampling the angle of rotation uniformly from the interval \([0, 2\pi)\). We then find the optimal Euler angles \((\beta', \gamma', \delta')\) via gradient descent of Eq. \eqref{eqn:EF_theta_max} over the parameters \((\beta, \gamma, \delta)\). Finally, we sample the initial state \(\ket{\psi(\theta, \phi)}\) uniformly from the portion of the Bloch sphere with polar angle less than \(\theta\textsubscript{max}\), simulate the application of \(U(\beta', \gamma', \delta')\) and \(U(\beta\textsubscript{t}, \gamma\textsubscript{t}, \delta\textsubscript{t})\) in the presence of noise parameterized by \(\lambda\), and calculate the increase in fidelity to the target state \(U(\beta\textsubscript{t}, \gamma\textsubscript{t}, \delta\textsubscript{t})\ket{\psi(\theta, \phi)}\). We note that this represents the increase in expected fidelity due to the random sampling of the initial state. We repeat this for 100 randomized rotations and calculate the average increase in expected fidelity. The results over a range of possible noise levels \(\lambda\) and polar angles \(\theta\textsubscript{max}\) are displayed in Fig. \ref{fig:knowledge_noise_plot}. For \(\theta\textsubscript{max} = 0\), we have perfect knowledge of the initial state \(\ket{0}\) and accordingly recover Fig. \ref{fig:avg_prep_fidelity_plot}. For \(\theta\textsubscript{max} = \pi\), we have no knowledge of the initial state and accordingly see no improvement in fidelity. Intuitively, the more we know about the initial state of the input qubit, the more we can fine-tune our optimization of the target operation to achieve a higher expected fidelity.

\subsection*{Sensitivity to Coherence Time Drifts on \texttt{ibmq\_bogota}}\label{experiment_2}
    The optimization protocol outlined in this paper reduces the error rate of a single-qubit gate by employing the \(T_1\) and \(T_2\) times characterizing the qubit's decoherence. These noise parameters are obtained empirically and are therefore subject to error as a result of coherence time drifts between measurements (see Fig.~\ref{fig:bogota_drifts}). In the RB experiment outlined below, we demonstrate that our technique can improve single-qubit gate fidelity even when the assumed noise parameters are reasonably inaccurate.
    
    In reality, the coherence times assumed by the optimizer are fixed, and the system's coherence times are free to drift. However, since the system's coherence times are not controllable parameters, we simulate the effect of a drift by intentionally providing the optimizer with inaccurate \(T_{1,2}\) times. By varying the assumed \(T_{1,2}\) times and measuring the fidelity of the optimized decomposition, we quantify the sensitivity of the optimizer to inaccurate coherence times. Though \(T_1\) and \(T_2\) can generally drift by different factors, we simplify our experiment by simulating drifts of \(T_1\) and \(T_2\) from their assumed values by the same coherence time drift factor \(k\):
    \begin{equation}
        (\text{system } T_{1,2}) = k \times (\text{assumed } T_{1,2})\,.
    \end{equation}
    We justify this simplification in two ways. First, with limited reservable time on \texttt{ibmq_bogota} and limited computational resources for optimization, the experiment described in this section could only be made computationally feasible by reducing the scan of possible drifts from 2D to 1D. Second, though \(T_1\) and \(T_2\) do not generally drift by a common factor, they are indeed correlated, since \(T_2 \leq 2 \cdot T_1\). We note that in many systems, \(T_2  \ll T_1\), and \(T_2\) is thus only weakly correlated with \(T_1\). Nonetheless, up to one week before and after \texttt{ibmq_bogota} qubit 2 was used in this study, \(T_1\) and \(T_2\) were moderately correlated with a correlation coefficient of 0.68 (see Fig. \ref{fig:bogota_drifts}). Albeit an approximation, the dimensionality reduction that made this experiment computationally feasible is thus statistically motivated. For each scaled pair of coherence times assumed by the optimizer, we perform the same RB experiment used to generate Fig. \ref{fig:experiment_1_plot}, now optimizing each gate in a 300-gate circuit and measuring the fidelity at depths of 100, 200, and 300 gates. The fidelities are once again averaged over 10 randomized gate sequences.
    
\begin{figure*}[ht]
        \centering
       \includegraphics[width=2.0\columnwidth]{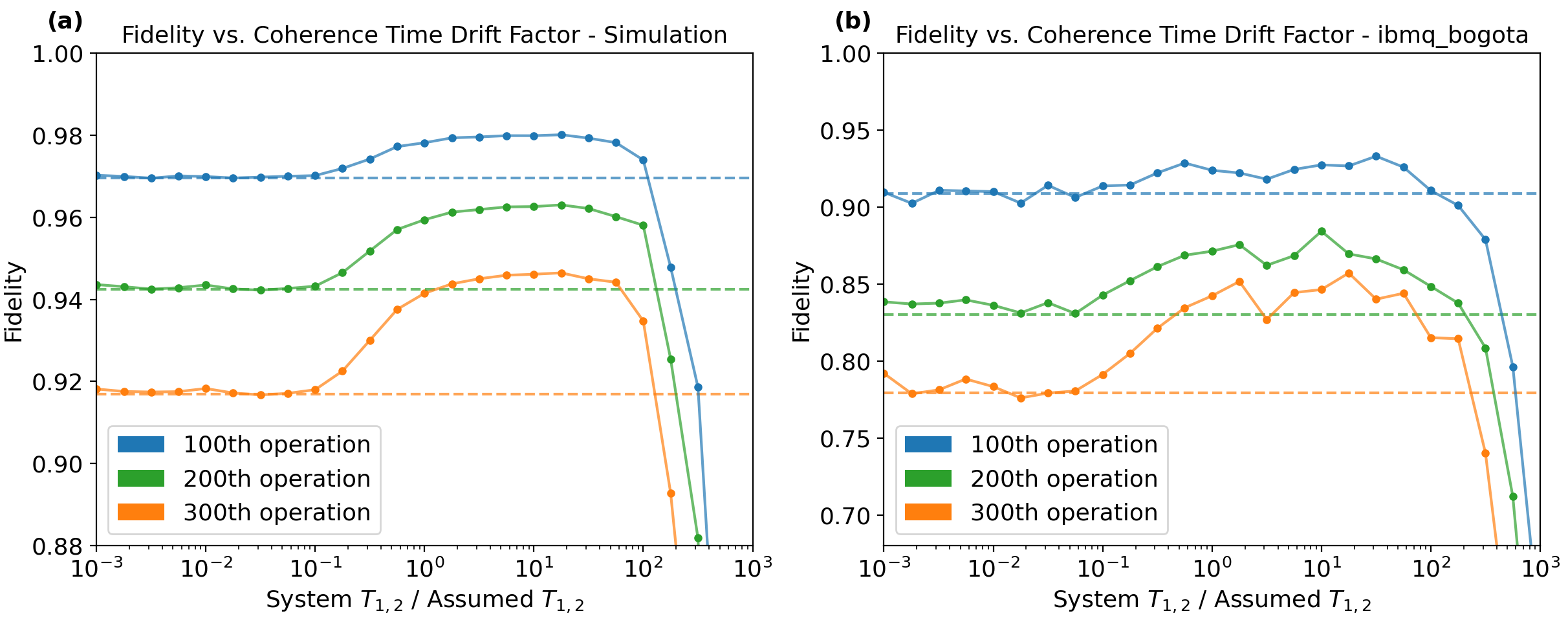}
        \caption{Fidelity vs. Coherence Time Drift Factor (System \(T_{1,2}\) / Assumed \(T_{1,2}\)): the fidelity (vertical axis) representing the overlap between the state output by the noisy application of the optimized circuit and the target state output by the noiseless application of the unoptimized circuit is plotted against the factor (horizontal axis) by which the system \(T_{1,2}\) times differ from their assumed values during optimization. Horizontal dashed lines represent represent unoptimized fidelities. Each data point is the average fidelity of 10 randomized gate sequences with 16,384 shots per measurement. Measurements were taken at circuit depths increasing by 100, \(d \in \{100, 200, 300\}\).}
        \label{fig:slice_plot}
    \end{figure*}

    The results from simulating the execution of the circuits using our noise model and from executing the circuits on \texttt{ibmq\_bogota} qubit 2 are displayed in Fig. \ref{fig:slice_plot}(a) and Fig. \ref{fig:slice_plot}(b), respectively. Each color corresponds to a particular circuit depth, as indicated by the legend. For each circuit depth, the corresponding dashed line represents the fidelity obtained without any optimization. The accumulation of noise in all circuits is reflected in the decrease in fidelity with circuit depth. As with the first experiment, the fidelities observed on the hardware are lower than those predicted by the simulation. We once again attribute this to noise channels unaccounted for by our noise model, as well as to the approximation that noisy single-qubit gates can be modeled as coherent evolution followed by decoherent evolution.
    
    As the factor \(k\) approaches 0, the optimizer assumes that there is no noise in the system. For any target rotation, in the absence of noise, the optimal rotation is the target rotation itself. Thus, as \(k\) approaches 0, we expect the fidelity offered by the optimizer to approach the unoptimized fidelity. This holds true in both the simulation data and the hardware data in Fig. \ref{fig:slice_plot}. On the other hand, as \(k\) approaches \(\infty\), the optimizer assumes an unrealistically large amount of noise in the system. The gradient descent is therefore performed over an assumed landscape unrepresentative of the system. The optimal rotation found by the optimizer will thus become uncorrelated with the target rotation, and the resulting fidelity will drop to 0.5 on average. This limit is also correctly captured in both the simulation data and the experiments performed on the quantum processor; however, since this regime is impractical, we have omitted much of this fidelity drop-off for the sake of visualization. In between the extremes, we expect to see a fidelity greater than the unoptimized fidelity. Once again, this is correctly reflected in both the simulation and experimental data. Most importantly, when the initial state of the input qubit is known, the optimizer will only decrease gate fidelity if the coherence times have drifted above about 100 times their assumed values. Since such drastic drifts are unrealistic (see Fig. \ref{fig:bogota_drifts}), we conclude that the optimizer is extremely unlikely to decrease gate fidelity and is therefore robust against coherence time drifts.

\section*{Discussion}\label{outlook}

    The optimization protocol outlined in this work reduces the error rate of single-qubit gates by leveraging knowledge of the initial state of the input qubit, along with the level of decoherence in the system, defined via \(T_1\) and \(T_2\) coherence times. On the \texttt{ibmq\_rome} quantum processor, we proved that the protocol can reduce single-qubit error rates by 38\%, from \(1.6 \times 10^{-3}\) to \(1.0 \times 10^{-3}\), provided the initial state of the input qubit is known. On \texttt{ibmq\_bogota}, we showed that the protocol always increases gate fidelity, provided the \(T_1\) and \(T_2\) times have not drifted below 0.1 or above 100 times their assumed values. The protocol can improve the expected fidelity of a single-qubit gate provided \emph{some} knowledge of the initial state of the input qubit—the more localized the initial state distribution, the more we can improve the expected fidelity. The technique can be applied as a means to improve the fidelity of state preparation, as well as to improve the fidelity of quantum circuits for which some knowledge of intermediate states of qubits can be inferred; the former is well-suited for use in variational quantum eigensolvers (VQE), while the latter is likely only applicable in the near-term. Our optimization technique is not limited to the native gate set of IBM quantum hardware and can be adapted for optimization over an arbitrary native gate set.
    
    In the noise model employed in this work, we used the approximation that native gates are applied instantaneously and are followed by a period of decoherence over the time interval equal to the gate duration. This approximation can be avoided by deriving a noise model from master equations involving the system's Hamiltonian during gate execution (e.g., the Linbladian for Markovian environments), which falls outside of the scope of this work. If the exact initial state of the input qubit is unknown, the optimization protocol can be modified to instead maximize the expected fidelity, averaged over the distribution of possible initial states. Our technique can also be adapted to optimize \(n\)-qubit gates by parameterizing all possible decompositions into an expanded native gate set (including entangling gates), and then proceeding as usual with the noise model and expected fidelity maximization for a specified distribution of \(n\)-qubit initial states. Since this work focused specifically on the optimization of single-qubit gates, gradient descent was sufficient for fidelity maximization; however, since the degrees of freedom required to parameterize an \(n\)-qubit gate are exponential in \(n\), efficient alternatives to gradient descent are likely necessary to maximize the fidelity of larger \(n\)-qubit operations with more practical run-times.
    
   In the near-term, while the execution of quantum circuits can still be simulated by classical computers, one can track the state of each qubit and use our protocol to optimize an arbitrary gate embedded within a quantum circuit. While the target qubit to said gate will generally be entangled with other qubits in the circuit, its reduced density matrix can be recovered by tracing over the degrees of freedom introduced by the qubits with which it is entangled. Though this entanglement will cause the qubit's initial state to be mixed, the optimization protocol can be easily adapted to accommodate mixed initial states, as is alluded to in Appendix \ref{noise_model_derivation}. We emphasize that in the near-term, the protocol can be used to optimize all single-qubit gates on NISQ devices, beyond those involved in state preparation. This work presents a pathway to using information about noise levels and quantum state distributions to significantly reduce error rates associated with quantum gates via optimized decomposition into native gates.

\section*{Data Availability}
The datasets generated during and/or analyzed during the current study are available from the corresponding author on reasonable request.

\bibliography{refs}

\section*{Acknowledgements}
We acknowledge the use of IBM Quantum services for this work. The views expressed are those of the authors, and do not reflect the official policy or position of IBM or the IBM Quantum team. We similarly thank Rigetti for access to its Aspen-8 processor. We also acknowledge the use of Google Slides and the Quantum Toolbox in Python (QuTiP) for the generation of Fig. \ref{fig:noisy_application} and Fig. \ref{fig:bloch_sphere}, respectively. The Flatiron Institute is a division of the Simons Foundation.

\section*{Author contributions statement}

T.M., J.F., S.K., and A.G. conceived the work, T.M. carried out the calculations, and all authors contributed to the interpretation of the results and the writing of the manuscript.

\section*{Competing interests}

The authors declare no competing interests.
\clearpage

\appendix

    \section{Noise Model Derivation} \label{noise_model_derivation}
        As discussed in the Noise Model section of the main text, we model the noisy application of an \(R_x(\pm\pi/2)\) gate as an instantaneous rotation, followed by decay and dephasing over time \(t_*\) equal to the gate duration. Employing IBM's publicly available \(T_1\) and \(T_2\) times, along with the gate duration \(t_*\), we define the following Kraus operators for the amplitude damping and phase damping noise channels, respectively:
        \begin{equation}
            \label{eqn:amp_damp_ops}
            A_0=
            \begin{bmatrix}
                1 & 0 \\
                0 & \sqrt{1-\lambda\textsubscript{A}}
            \end{bmatrix}
            \quad A_1=
            \begin{bmatrix}
                0 & \sqrt{\lambda\textsubscript{A}} \\
                0 & 0
            \end{bmatrix}
        \end{equation}
        \begin{equation}
            \label{eqn:phase_damp_ops}
            P_0=
            \begin{bmatrix}
                1 & 0 \\ 
                0 & \sqrt{1-\lambda\textsubscript{P}}
            \end{bmatrix}
            \quad P_1=
            \begin{bmatrix}
                0 & 0 \\ 
                0 & \sqrt{\lambda\textsubscript{P}}
            \end{bmatrix}
        \end{equation}
       The variables \(\lambda\textsubscript{A}\) and \(\lambda\textsubscript{P}\) are defined in Eqs.~\eqref{eqn:lambda_A}-\eqref{eqn:lambda_P}~of the main text, and their physical significance is described in the Noise Model section of the main text. Since the amplitude damping and phase damping superoperators commute, we can unambiguously define a mapping \(N\) that encapsulates the net effect of the two noise channels by applying one after the other, thereby mapping the density matrix
        \begin{equation}
            \label{eqn:rho}
            \rho = \begin{bmatrix}
                \rho_{00} & \rho_{01} \\
                \rho_{10} & \rho_{11}
            \end{bmatrix}
        \end{equation}
        to the density matrix 
        \begin{equation}
            \label{eqn:N}
            N(\rho)= 
                \begin{bmatrix}
                    \rho_{00}(1 - \lambda\textsubscript{A}) + \lambda\textsubscript{A} & \rho_{01}\sqrt{1 - \lambda\textsubscript{A}}\sqrt{1 - \lambda\textsubscript{P}} \\
                    \rho_{10}\sqrt{1 - \lambda\textsubscript{A}}\sqrt{1 - \lambda\textsubscript{P}} & \rho_{11}(1 - \lambda\textsubscript{A})
                \end{bmatrix}
        \end{equation}
        Using this mapping and the decomposition defined by Eq.~\eqref{eqn:ibm_decomp}~of the main text, we proceed by explicitly calculating the density matrix of a qubit after the noisy application of a single-qubit gate with Euler angles \((\beta, \gamma, \delta)\). If needed, we could begin with an initially mixed state parameterized by a 3-dimensional Bloch vector with length no more than unity, but for the sake of simplicity, we assume the input qubit is in an initially pure state with Bloch sphere coordinates \((\theta, \phi)\). The initial state vector is given by the following:
        \begin{equation}
            \label{eqn:input_state_vector}
            \ket{\psi(\theta, \phi)} = \cos{(\theta/2)}\ket{0} + e^{i\phi}\sin{(\theta/2)}\ket{1}
        \end{equation}
        The qubit's density matrix after the first pulse is given by:
        \begin{equation}
            \rho_1 = N(R_x(\pi/2)R_z(\delta)\ket{\psi(\theta, \phi)}\bra{\psi(\theta, \phi)}R_z(\delta)^{\dagger}R_x(\pi/2)^{\dagger})
        \end{equation}
        After the second pulse:
        \begin{equation}
            \rho_2 = N(R_x(-\pi/2)R_z(\gamma)\rho_1R_z(\gamma)^{\dagger}R_x(-\pi/2)^{\dagger})
        \end{equation}
        After the final \(R_z(\beta)\) rotation:
        \begin{equation}
            \rho_{(\beta, \gamma, \delta, \theta, \phi)} = R_z(\beta)\rho_2R_z(\beta)^{\dagger}
        \end{equation}
        Applying these transformations, we find that the noisy application of a single-qubit gate parameterized by Euler angles \((\beta, \gamma, \delta)\) transforms an initially pure state with Bloch sphere coordinates \((\theta, \phi)\) into a mixed state with the density matrix defined in Eqs.~\eqref{eqn:noisy_density_matrix}-\eqref{eqn:noisy_density_matrix_b}~of the main text.
        
        Throughout this derivation, we have implicitly assumed that the calibrated pulse amplitudes used to implement the \(R_x(\pm\pi/2)\) gates in the presence of noise were exactly equal to the pulse amplitudes that would implement the \(R_x(\pm\pi/2)\) gates perfectly in the absence of noise. This assumption is consistent with IBM's calibration methodology. IBM calibrates their \(R_x(\pm\pi/2)\) gates by performing a Rabi experiment to determine the pulse amplitudes that most accurately implement them in the presence of noise \cite{Ashhab_2006, Galperin_2005, Majer_2007}. Concretely, the optimal pulse is that which most accurately maps the \(\ket{0}\) state to the state \(R_x(\pm\pi/2)\ket{0}\) in the presence of noise. Within the framework of our noise model, since the angle of an \(R_x\) rotation is proportional to the amplitude of the pulse that implements it, finding the optimal pulse amplitude amounts to finding the optimal angle \(\alpha\) that maximizes the following fidelity:
        \begin{align}
            \label{eqn:calibration_fidelity}
            \begin{split}
                F &= \bra{0}R_x(\pm\pi/2)^\dagger N(R_x(\alpha)\ket{0}\bra{0}R_x(\alpha)^\dagger) R_x(\pm\pi/2)\ket{0} \\
                &= \frac{1}{2}(1 \pm \sqrt{1-\lambda\textsubscript{A}}\sqrt{1-\lambda\textsubscript{P}}\sin{\alpha})
            \end{split}
        \end{align}
        Since this achieves a maximum at \(\alpha = \pm \pi/2\), it follows that the calibrated pulse amplitudes used to implement the \(R_x(\pm\pi/2)\) gates in the presence of noise are exactly equal to the pulse amplitudes that would implement the \(R_x(\pm\pi/2)\) gates perfectly in the absence of noise.
        
    \section{Error Rate Reduction on Aspen-8}\label{experiment_3}
        \begin{figure*}
            \centering
            \includegraphics[width=2.0\columnwidth]{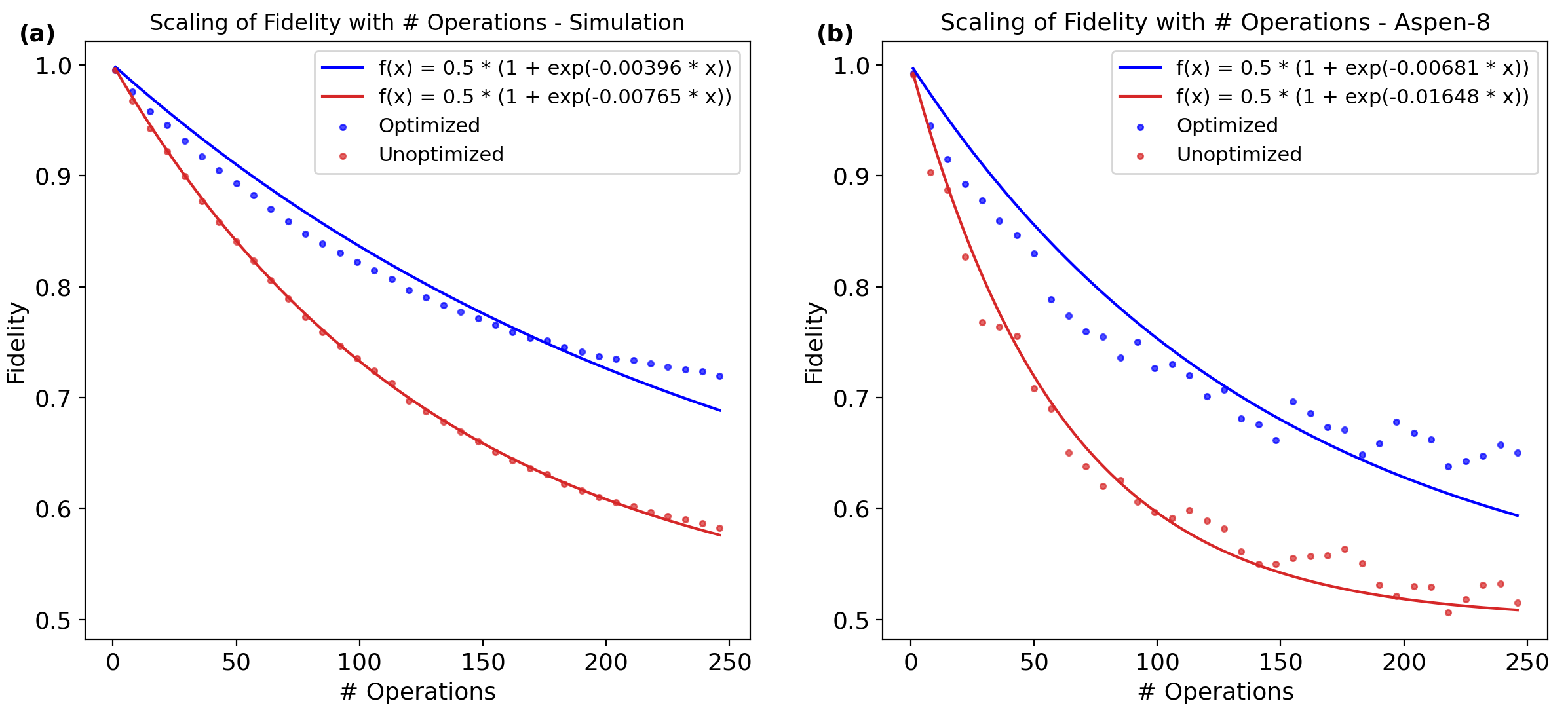}
            \caption{Scaling of fidelity with number of operations: the fidelity (vertical axis) representing the overlap between the state output by the noisy application of each circuit (unoptimized and optimized) and the target state output by the noiseless application of the unoptimized circuit is plotted against the circuit depth (horizontal axis) at which the fidelity was measured. Each data point is the average fidelity of 10 randomized gate sequences with 4,096 shots per measurement. Measurements were taken at circuit depths increasing by 7, \(d \in \{1, 8, 15, ..., 246\}\).}

            \label{fig:rigetti_scaling}
        \end{figure*}
        Since Rigetti's QCs are built from superconducting transmon qubits, their native gate set is identical to IBM's, allowing for easy adaptation of our protocol. On Rigetti's Aspen-8 qubit 4 \cite{rigetti}, we performed the same experiment outlined in the ``Error Rate Reduction on \texttt{ibmq\_rome}'' subsection of the main text. Readout error was mitigated for all measurements made on Aspen-8 by inverting a calibration matrix \cite{readouterror} composed of Rigetti's publicly reported readout error probabilities. All single-qubit noise specifications for Aspen-8 were provided by Rigetti through PyQuil \cite{rigetti, smith2016practical, Karalekas_2020} and are tabulated in the Noise Specifications section below. The results of the experiment are displayed in Fig. \ref{fig:rigetti_scaling}. An analysis of this data would mirror the analysis in the main text, so to avoid redundancy, we only discuss the reduction in error rate here.
        
        On Aspen-8 qubit 4, our optimization protocol reduces the error rate of a single-qubit gate acting on a known initial state by 59\%, from \(8.2 \times 10^{-3}\) to \(3.4 \times 10^{-3}\). The unoptimized error rate that we report here does not agree with Rigetti's reported error rate of \(7.5 \times 10^{-3}\), but there is better agreement than was found on \texttt{ibmq\_rome}.

    \section{Noise Specifications}\label{noise_specs}
        For all of the QCs used throughout this work, the single-qubit noise specifications on their respective dates of use are tabulated below.
        \noindent\\
        \begin{table*}
            \centering
            \begin{tabular}{|c|c|c|c|c|c|c|c|}
                \hline
                \multicolumn{7}{|c|}{\textbf{Noise Specifications - ibmq\_rome - 07/14/20}} \\
                \hline
                Qubit & \(T_1\) \([\mu s]\) & \(T_2\) \([\mu s]\) & Pulse Duration \([ns]\) & Prob. Prep. \(\ket{0}\) Meas. \(\ket{1}\) & Prob. Prep. \(\ket{1}\) Meas. \(\ket{0}\) & Gate Error [\(10^{-4}\)]\\
                \hline
                0 & 116 & 93.4 & 35.6 & 0.030 & 0.063 & 3.23\\
                \hline
                1 & 105 & 55.2 & 35.6 & 0.020 &	0.073 & 3.00\\
                \hline
                2 & 80.2 & 124 & 35.6 & 0.073 & 0.087 & 4.04\\
                \hline
                3 & 46.4 & 105 & 35.6 & 0.027 & 0.050 & 3.35\\
                \hline
                4 & 71.6 & 138 & 35.6 & 0.027 & 0.043 & 5.01\\
                \hline
                Mean & 83.9 & 103 & 35.6 & 0.035 & 	0.063 & 3.77\\
                \hline
                SD & 27.7 & 31.7 & 0.00 & 0.022 & 0.017 & 0.798 \\
                \hline
            \end{tabular}
            
            \centering
            \begin{tabular}{|c|c|c|c|c|c|c|c|}
                \hline
                \multicolumn{7}{|c|}{\textbf{Noise Specifications - ibmq\_bogota - 08/10/20}} \\
                \hline
                Qubit & \(T_1\) \([\mu s]\) & \(T_2\) \([\mu s]\) & Pulse Duration \([ns]\) & Prob. Prep. \(\ket{0}\) Meas. \(\ket{1}\) & Prob. Prep. \(\ket{1}\) Meas. \(\ket{0}\) & Gate Error [\(10^{-4}\)]\\
                \hline
                0 & 126 & 158 & 35.6 & 0.024 & 0.108 & 5.10\\
                \hline
                1 & 117 & 168 & 35.6 & 0.004 &	0.039 & 2.36\\
                \hline
                2 & 107 & 142 & 35.6 & 0.009 & 0.026 & 2.19\\
                \hline
                3 & 199 & 240 & 35.6 & 0.005 & 0.034 & 3.96\\
                \hline
                4 & 142 & 249 & 35.6 & 0.019 & 0.135 & 2.02\\
                \hline
                Mean & 138 & 192 & 35.6 & 0.012 & 0.068 & 2.63\\
                \hline
                SD & 36.4 & 49.6 & 0.00 & 0.009 & 0.050 & 0.898 \\
                \hline
            \end{tabular}
            \begin{tabular}{|c|c|c|c|c|c|c|c|}
                \hline
                \multicolumn{7}{|c|}{\textbf{Noise Specifications - Aspen-8 - 08/30/20}} \\
                \hline
                Qubit & \(T_1\) \([\mu s]\) & \(T_2\) \([\mu s]\) & Pulse Duration \([ns]\) & Prob. Prep. \(\ket{0}\) Meas. \(\ket{1}\) & Prob. Prep. \(\ket{1}\) Meas. \(\ket{0}\) & Gate Error [\(10^{-4}\)]\\
                \hline
                0 & 16.9 & 18.7 & 60.0 & 0.014 & 0.034 & 50.0\\
                \hline
                1 & 36.7 & 40.4 & 60.0 & 0.049 & 0.063 & 4.00\\
                \hline
                2 & 18.5 & 12.1 & 60.0 & 0.017 & 0.033 & 112\\
                \hline
                3 & 34.2 & 19.5 & 60.0 & 0.022 & 0.046 & 13.0\\
                \hline
                4 & 15.3 & 17.6 & 60.0 & 0.017 & 0.034 & 75.0\\
                \hline
                5 & 41.4 & 5.37 & 60.0 & 0.047 & 0.075 & 5.00\\
                \hline
                6 & 45.0 & 45.1 & 60.0 & 0.015 & 0.042 & 31.0\\
                \hline
                7 & 17.2 & 22.8 & 60.0 & 0.016 & 0.038 & 11.0\\
                \hline
                11 & 23.8 & 12.7 & 60.0 & 0.058 & 0.058 & 10.0\\
                \hline
                12 & 21.6 & 8.18 & 60.0 & 0.034 & 0.134 & 27.0\\
                \hline
                13 & 27.0 & 18.3 & 60.0 & 0.040 & 0.056 & 26.0\\
                \hline
                14 & 17.9 & 11.3 & 60.0 & 0.018 & 0.037 & 40.0\\
                \hline
                15 & 35.9 & 3.76 & 60.0 & 0.035 & 0.066 & 16.0\\
                \hline
                16 & 21.8 & 28.6 & 60.0 & 0.009 & 0.037 & 24.0\\
                \hline
                17 & 37.9 & 14.7 & 60.0 & 0.021 & 0.042 & 13.0\\
                \hline
                20 & 18.7 & 15.7 & 60.0 & 0.017 & 0.041 & 19.0\\
                \hline
                21 & 43.7 & 7.29 & 60.0 & 0.045 & 0.082 & 32.0\\
                \hline
                22 & 29.6 & 25.3 & 60.0 & 0.035 & 0.072 & 19.0\\
                \hline
                23 & 24.8 & 9.89 & 60.0 & 0.074 & 0.078 & 1034\\
                \hline
                24 & 12.9 & 2.18 & 60.0 & 0.042 & 0.116 & 18.0\\
                \hline
                25 & 42.9 & 20.4 & 60.0 & 0.035 & 0.060 & 10.0\\
                \hline
                26 & 10.6 & 2.22 & 60.0 & 0.015 & 0.047 & 79.0\\
                \hline
                27 & 43.1 & 18.7 & 60.0 & 0.036 & 0.074 & 8.00\\
                \hline
                30 & 21.1 & 26.3 & 60.0 & 0.040 & 0.178 & 70.0\\
                \hline
                31 & 42.0 & 37.2 & 60.0 & 0.035 & 0.092 & 5.00\\
                \hline
                32 & 43.1 & 56.8 & 60.0 & 0.022 & 0.076 & 10.0\\
                \hline
                33 & 29.0 & 27.0 & 60.0 & 0.027 & 0.080 & 9.00\\
                \hline
                34 & 17.9 & 21.0 & 60.0 & 0.010 & 0.034 & 15.0\\
                \hline
                35 & 30.5 & 35.3 & 60.0 & 0.052 & 0.091 & 2.00\\
                \hline
                36 & 33.8 & 23.4 & 60.0 & 0.041 & 0.091 & 45.0\\
                \hline
                37 & 35.3 & 22.8 & 60.0 & 0.041 & 0.053 & 10.0\\
                \hline
                Mean & 28.7 & 20.3 & 60.0 & 0.032 & 0.066 & 59.4\\
                \hline
                SD & 10.7 & 12.8 & 0.00 & 0.016 & 0.033 & 183 \\
                \hline
            \end{tabular}
        \end{table*}

\renewcommand{\labelenumi}{S\arabic{enumi}}

\end{document}